\documentclass[letterpaper,aps,prl,twocolumn,amsmath,showpacs,amssymb]{revtex4}

\usepackage{epsfig}
\usepackage{graphicx}
\usepackage{dcolumn}
\usepackage{bm}
\usepackage{bbm}
\usepackage{wasysym}
\usepackage{marvosym}
\usepackage{times,amsmath,amssymb}

\newlength{\textwidthm}

\begin{document}

\author{N.~M.~R. Peres$^1$, F.~D. Klironomos$^2$, 
S.-W. Tsai$^2$, J.~R. Santos $^1$,
J.~M.~B. Lopes dos Santos$^3$, and A.~H. Castro Neto$^4$}

\affiliation{$^1$Center of Physics and Departamento de
F\'{\i}sica, Universidade do Minho, P-4710-057, Braga, Portugal}

\affiliation{$^2$Department of Physics and Astronomy, University of California, 
Riverside, CA 92521, USA}

\affiliation{$^3$CFP and Departamento de F{\'\i}sica, Faculdade de Ci\^encias
Universidade de Porto, 4169-007 Porto, Portugal}

\affiliation{$^4$Department of Physics, Boston University, 590 
Commonwealth Avenue,
Boston, MA 02215,USA}

\title{Electron waves in chemically substituted graphene}

\date{\today}

\begin{abstract}
We present exact analytical and numerical results for the electronic spectra 
and the Friedel oscillations around a substitutional impurity atom in a 
graphene lattice. A chemical dopant in graphene introduces changes in the 
on-site potential as well as in the hopping amplitude. We employ a $T$-matrix 
formalism and find that disorder in the hopping introduces additional
 interference terms around the impurity that can be understood in terms 
of bound, semi-bound, and unbound processes for the Dirac electrons. 
These interference effects can be detected by scanning tunneling microscopy.
\end{abstract}


\maketitle

\begin{figure}[htf]
\begin{center}
\includegraphics*[angle=-90,totalheight=6.5cm,width=7.5cm,viewport=5 5 560 720,
clip]{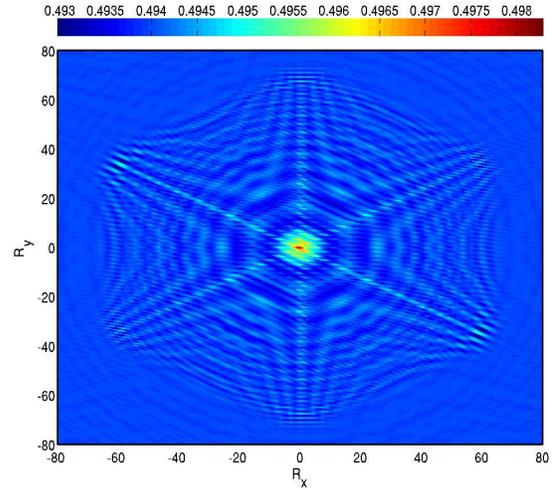}
\end{center}
\caption{(color on line) Real-space electronic distribution and Friedel oscillations 
around an impurity with ionic radius 
smaller than carbon (such as N) ($t_0=0.5t$).}
\label{fig:nr05}
\end{figure}

{\it Introduction.} Graphene \cite{novo1}, a one atom-thick layer of graphite, has been intensively studied 
\cite{novo2,novo3,kim1,kim2,kim3,berger,peres1,world,geim,pereira,loktev} 
due to its  fascinating physical properties \cite{peres1,world,geim}.   
Graphene is a zero-gap semiconductor and its low-energy electronic excitations are described in terms of a 
Dirac spectrum. Because of this, disorder in the form of impurities \cite{peres1,pereira,loktev,cheianov,wehling}, 
defects \cite{voz1,voz2,yazyev1,yazyev2} and surfaces \cite{ahcn1,peres2} can have a strong effect, especially when 
the chemical potential crosses the Dirac point. In this work, we consider  dilute chemical dopants in graphene 
incorporated as substitutional atoms, and calculate local single particle properties such as the electronic energy spectra, 
local density of states, and electron density distribution, which shows Friedel oscillations as depicted in Fig.~(\ref{fig:nr05}).
The results presented here can be measured by scanning tunneling microscopy (STM) \cite{stm1,stm2,eandrei}.

Atomic substitution in a carbon (C) honeycomb lattice is chemically possible for boron (B) and nitrogen (N) atoms. 
There have been several experimental studies of B and N substitution in highly-oriented pyrolytic graphite (HOPG) \cite{mele,endo}, 
graphitic structures \cite{stephan}, nanoribbons \cite{ze}, carbon nanotubes 
\cite{stephan, nanotube1,nanotube2,nanotube3,nanotube4,nanotube5,nanotube6,nanotube7,nanotube8,nanotube9,nanotube10} 
and fullerenes \cite{bucky}. When B or/and N atoms replace a carbon atom in a graphene sheet they have the following effects: 
(1) they act as impurity scattering centers; (2) they act as hole- or electron-dopants; (3) they introduce lattice distortions. 
In the case of B there is an increase of the absolute value of the hopping
amplitude given its larger atomic radius  ($r\simeq 0.85$\AA) 
as compared to C ($r\simeq 0.7$\AA). On the other hand, a N atom  
impurity ($r\simeq 0.65$\AA)
can be modeled by a smaller hopping integral. For the change in the on-site
energy, we assume the C on-site energy to be zero (our reference state);
the smaller atomic number of B leads to a positive local
energy, whereas on a site occupied by N, the local 
energy should be smaller and therefore is modeled as a negative
local energy. The effect of a different on-site energy has been
extensively studied in the past \cite{peres1,pereira,loktev,cheianov,wehling}, but the effect on the hopping has been overlooked.
In this work we study the combined effect
of local changes in the atomic energy as well as the hopping amplitudes.
 A vacancy has been modeled by an infinite,
on-site energy potential, but it is also exactly represented
by the remotion of particular hopping processes from the Hamiltonian. 
The opposite limit of very large hopping to the impurity site corresponds 
to a four-atom molecule. 
  
\begin{figure}[htf]
\begin{center}
\includegraphics*[totalheight=4.5cm,width=5.5cm]{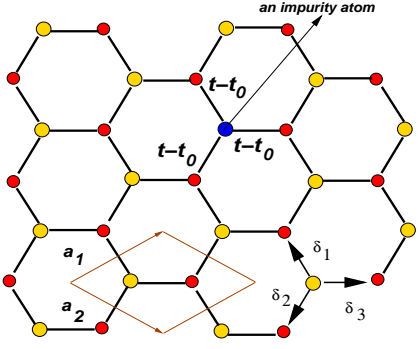}
\end{center}
\caption{(color on line)
\label{Fig_lattice} The honeycomb lattice with a carbon atom replaced by
an impurity atom, such a B or N. The local hopping parameter changes from 
$t$ to $t-t_0$.}
\end{figure}

 {\it Hopping and potential disorder.} The honeycomb lattice has a unit cell 
represented in Fig.~(\ref{Fig_lattice})
by the vectors $\bm a_1$ and $\bm a_2$, such that 
$\vert \bm a_1 \vert=\vert \bm a_2 \vert=a$ ($a = \sqrt{3} a_0 \simeq 2.461$ \AA, where $a_0$
is the carbon-carbon distance). In this basis any 
lattice vector $\bm r$
is represented as
$\bm r= n\bm a_1 + m\bm a_2$, with $n,m$ integers. 
In Cartesian coordinates,
$\bm a_1 = a_0 (3,\sqrt 3,0)/2$ and 
$\bm a_2 = a_0 (3,-\sqrt 3,0)/2$.
The reciprocal lattice vectors are given by:
$\bm b_1= 2\pi (1,\sqrt 3,0)/(3 a_0)$ and 
$\bm b_2= 2\pi (1,-\sqrt 3,0)/(3 a_0)$,
and the  vectors connecting any $A$ atom to its nearest
neighbors are: 
$\bm \delta_1 = (\bm a_1-2\bm a_2)/3$, 
$\bm \delta_2 = (\bm a_2-2\bm a_1)/3$, and 
$\bm \delta_3 = (\bm a_1+\bm a_2)/3$.
Using these 
definitions the Hamiltonian can be written as: $H=H_0+V_t+V_0$, where 
$
H_0 = -t \sum_{\bm r} [b^\dag(\bm r) a(\bm r)+ b^\dag (\bm r - \bm a_2) a(\bm r)
+ b^\dag(\bm r - \bm a_1) a(\bm r) + {\rm h.c.}],
$
is the kinetic energy operator and $a^\dag$ ($b^\dag$) are creation
operators in the $A$ ($B$) sites.
The operators 
$V_t = t_0[b^\dag(0) a(0) + b^\dag(-\bm a_2)a(0) + b^\dag(-\bm a_1) a(0) + {\rm h.c.}]$ and 
$V_0=\varepsilon_0 a^{\dagger}(0)a(0)$ are the impurity terms for hopping and potential 
disorder, respectively. In the particular case $t_0=t$, the scattering term
$V_t$ represents a vacancy. 
In this work we consider the case of zero chemical potential when the Fermi level crosses the Dirac point. 
This is the case in which the system is most susceptible to the presence of impurities.
The equations of motion for the Green's functions can be readily written,
and are given by:
\begin{eqnarray}
\!\!i\omega_n G_{aa}\!(\omega_n,\!\bm k,\!\bm p\!)\!\!\!&=&\!\!\!\delta_{\bm k,\bm p} \!\!+\!\!\! \sum_{\bm q} \!\!\left[\! \lambda_{\bm k, \bm q} G_{ba}\!(\omega_n,\!\bm q,\! \bm p\!) \!+\!\! \frac{\varepsilon_0}{N_c}\!G_{aa}\!(\omega_n,\!\bm q,\!\bm p\!)\!\right]\nonumber \\
\!\!i\omega_n G_{ba}\!(\omega_n,\!\bm k,\!\bm p\!)\!\!\!&=&\!\!\! \sum_{\bm q} 
\lambda^{\ast}_{\bm q, \bm k} G_{aa}(\omega_n,\bm q,\bm p\!)\nonumber\\
\!\!i\omega_n G_{ab}\!(\omega_n,\!\bm k,\!\bm p\!)\!\!\!&=&\!\!\! \sum_{\bm q} \!\!\left[\!\lambda_{\bm k,\bm q} G_{bb}(\omega_n,\bm q,\bm p\!)\!+\!\! \frac{\varepsilon_0}{N_c}\!G_{ab}\!(\omega_n,\!\bm q,\!\bm p\!)\!\right] \nonumber\\
\!\!i\omega_n G_{bb}\!(\omega_n,\!\bm k,\!\bm p\!)\!\!\!&=&\!\!\! \delta_{\bm
  k,\bm p} \!+\! \sum_{\bm q} \lambda^{\ast}_{\bm q,\bm k} G_{ab}(\omega_n,\bm q,\bm
p\!) \, ,
\nonumber 
\end{eqnarray}
where $\lambda_{\bm k,\bm p} = -t\phi_{\bm p} (\delta_{\bm k, \bm p} - t_0/N_c t)$,
$\phi_{\bm p} =1+e^{-i\bm p\cdot \bm a_1}+e^{-i\bm p\cdot \bm a_2}$, and $N_c$
is the total number of unit cells in the lattice. 
The sublattice symmetry is broken by the presence of the impurity and this is 
manifested in the fact that $\lambda_{\bm p\bm q}\ne \lambda_{\bm q\bm p}$
and in the  asymmetric way
in which the $\varepsilon_0$-term appears in the equations above. 
The above set of equations can be solved exactly. The fact that the
scattering term $V_t$ depends on $\phi_{\bm k}$ leads to a more complex form 
for the $T$-matrix than usual. 
The exact solution for the Green's functions can be written in the form:
\begin{eqnarray}  
G_{aa}(\bm k,\bm p) \!&=&\! \delta_{\bm k,\bm p} G^0_{\bm k} + g +
   h \left[ G^0_{\bm k} + G^0_{\bm p} \right] 
+ G^0_{\bm k} T G^0_{\bm p} \;,
\label{gaa}\\
  G_{bb}(\bm k,\bm p) \!&=&\! \delta_{\bm k,\bm p} G^0_{\bm k} +
    \frac{t \phi^*_{\bm k}}{i\omega_n} G^0_{\bm k} T
    G^0_{\bm p} \frac{t \phi_{\bm p}}{i\omega_n} \;.
\label{gbb}
\end{eqnarray}
where all the terms also depend on $\omega_n$ (omitted here for brevity). 
They are defined as: 
\begin{eqnarray} 
g(\omega_n)&=& t_0^2 \bar{G}^0(\omega_n)/[N_c D(\omega_n)] , \\
h(\omega_n) &=&  t_0(t-t_0)/[N_c D(\omega_n)]  ,
\end{eqnarray}
and 
\begin{eqnarray}
T(\omega_n) =
   -\frac{i\omega_n t_0 (2t-t_0) - \varepsilon_0 t^2}{N_c D(\omega_n)} 
\end{eqnarray}
where the denominator $D(\omega_n)$ is given by
\begin{eqnarray}
  D(\omega_n) \!=\! (t\!-\!t_0)^2 \!+\!
  \left[ i\omega_n t_0 (2t\!-\!t_0) \!-\! \varepsilon_0 t^2 \right]\!
  \bar{G}^0(\omega_n)
\end{eqnarray}
and $\bar{G}^0(\omega_n) = \sum_{\bm k} G^0(\omega_n, \bm k)/N_c$ with 
$G^0_{\bm k}=  G^0(\omega_n, \bm k) =  (i\omega_n)/[(i\omega_n)^2-t^2|\phi_{\bm k}|^2],$ the diagonal Green's function for the clean system.

The significance of the term $g(\omega_n)$ in (\ref{gaa}) which only 
appears in  $G_{aa}$, 
is easily appreciated if we do the double Fourier transform to real space. 
This term only contributes to $G_{aa}(0,0)$, the return amplitude 
to the impurity site.
The factor 
$1/D(\omega_n)$ contains a sum over an infinite series of intermediate 
scattering events, but the overall process is bounded and the $t_0^2$ factor 
denotes hopping from the impurity to the nearest neighbor B-sites and back to the impurity site.  
A similar interpretation can be given to another term which only appears in
$G_{aa}$, namely, $G^{0}(\omega_n,k)h(\omega_n)$: this term contributes only to
$G_{aa}(r,0)$ and describes an additional amplitude of propagation between the 
impurity site and another $A$ site. No such terms can, of course, 
appear in $G_{bb}$
when the inpurity is at an $A$ site.  

The exact general solution for both diagonal components of the Green's functions for a substitutional impurity with potential and hopping disorder is contained in 
(\ref{gaa}) and (\ref{gbb}) .
The local density of states (LDOS) can be obtained from the Green's functions (\ref{gaa},\ref{gbb}), after analytical continuation 
$i\omega_n \rightarrow \omega + i0^+$. For sites in the $A$ and $B$ sublattices, it is given by:
$
\rho_{\nu}(\bm r,\omega) = - {\rm Im} G_{\nu\nu}(\omega,\bm r,\bm r)/\pi \, ,
$
where $\nu=a,b$ and $\bm r$ is the position of the unit cell. 
The local number of electrons, for a half-filled band, is obtained by:
\begin{eqnarray}
n_{a,b}(\bm r) = \int_{-W}^0 d\omega \rho_{a,b}(\bm r,\omega) \, ,
\label{friedel}
\end{eqnarray}
where $W=3t$ is half of the total electronic bandwidth.

\begin{figure}[thf]
\begin{center}
\includegraphics*[totalheight=4.8cm,width=6.2cm,viewport=110 65 1505 1190,clip]
{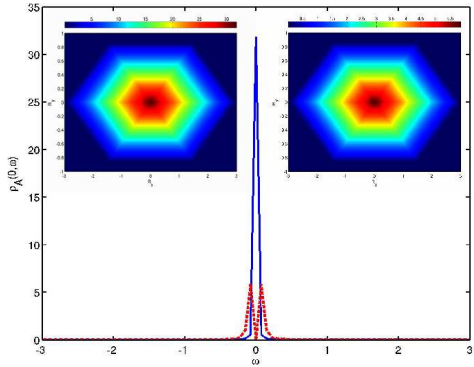}
\end{center}
\caption{(Color online) LDOS for $t_0=t$ and $t_0=0.9t$ at the impurity site.
Main graph corresponds to $\rho_a(0,\omega)$ for $t_0=t$ (straight blue
line) and for $t_0=0.9t$ (dashed red line). 
Left inset: $\rho_a(\bm r,\omega=0)$ for $t_0=t$. 
Right inset: $\rho_a(\bm r,\omega=-0.075t)$ for $t_0=0.9t$. Notice the scale
difference.}
\label{fig:A1_09}
\end{figure}

\begin{figure}[thf]
\begin{center}
\includegraphics*[totalheight=4.8cm,width=6.2cm,viewport=100 65 1505 1190,clip]
{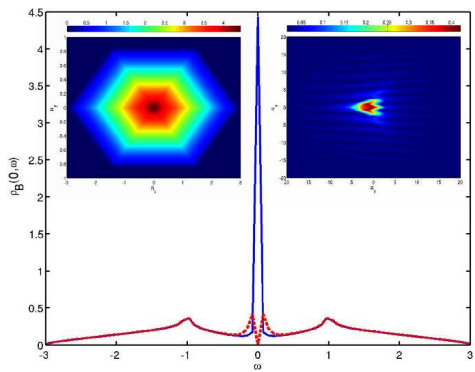}
\end{center}
\caption{(Color online) LDOS for $t_0=t$ and $t_0=0.9t$ at the nearest neighbor
sites of the impurity. Main graph corresponds to $\rho_b(0,\omega)$ for
$t_0=t$ (straight blue line) and for $t_0=0.9t$ (dashed red line).
Left inset: $\rho_b(\bm r,\omega=0)$ for $t_0=t$. 
Right inset: $\rho_b(\bm r,\omega=-0.075t)$ for $t_0=0.9t$.}
\label{fig:B1_09}
\end{figure}

{\it Strong suppression of hopping and vacancies.}
A vacancy corresponds to $t_0=t$, leading to a simplified
form for  $G_{aa}$:
\begin{eqnarray}
G_{aa}(\bm k,\bm p) =  G^0_{\bm k}\delta_{\bm k,\bm p}+
\frac 1 {N_c i\omega_n}
-\frac{G^0_{\bm k}[\bar{G}^0]^{-1}G^0_{\bm p}}{N_c}\, ,
\label{gaatot}
\end{eqnarray}
where $G_{aa}$, $G^0$ and $\bar{G}^0$ also depend on $\omega_n$, and the on-site impurity potential was set to zero ($\varepsilon_0=0$).
Because the vacancy creates a three-site zig-zag edge, one expects
the appearance of zero energy modes \cite{peres1}.  
The $1/i\omega_n$ term in (\ref{gaatot}) leads to a contribution, 
$\rho_{a}(0,\omega) \propto \delta(\omega)$ in the total DOS, as shown
in Fig.~(\ref{fig:A1_09}).
This contribution comes from the atom that has been disconnected from the rest
of the lattice. It corresponds to exactly one state and to properly represent a
missing atom this contribution to the DOS should be ignored. 
At small energies ($\omega \ll t$) we obtain for the nearest neighbor B-sites
that $G_{bb}(0,0,\omega)\approx -1/(9t^2\bar{G}^0(\omega))$, in agreement with previous
studies \cite{peres1,balatsky,palee} for vacancy modeled as infinite on-site potential, $\varepsilon_0 = \infty$. In this limit,
$
\rho_b(0,\omega)\simeq(\sqrt{3}|\omega|/6\pi t^2)\{1+(36\pi^2/27)[\omega\ln|
\sqrt{3}\omega^2/6\pi t^2|]^{-2}\}
$
giving a resonance at $\omega=0$. The full numerical calculation for all
energies is shown in Fig.~(\ref{fig:A1_09}) for the impurity site and in Fig.~(\ref{fig:B1_09}) for the nearest neighbor sites, with the spatial dependence of the amplitude of the low-energy states shown in the insets.

For $t_0$ slightly smaller than $t$, hopping to the impurity site is strongly 
suppressed but not completely absent. 
Hybridization between the low energy states at the impurity site and at the nearest neighbors sites leads to splitting into two resonant states, one with energy shifted to a negative value and the other to a positive value, as shown in  Figs.~(\ref{fig:A1_09}-\ref{fig:B1_09}). 

{\it Nitrogen substitution.} As discussed previously, a N atom has a smaller 
atomic radius than a carbon atom, and this corresponds 
to a smaller hopping amplitude between the N impurity and the neighboring
carbon atoms. In addition, the larger atomic number of N atoms 
gives a negative on-site potential $\varepsilon_0$ with respect to the carbon sites. 
\begin{figure}[htf]
\begin{center}
{\includegraphics*[angle=-90,totalheight=2.5cm,width=3.5cm,viewport=10 5 550 695,
clip]{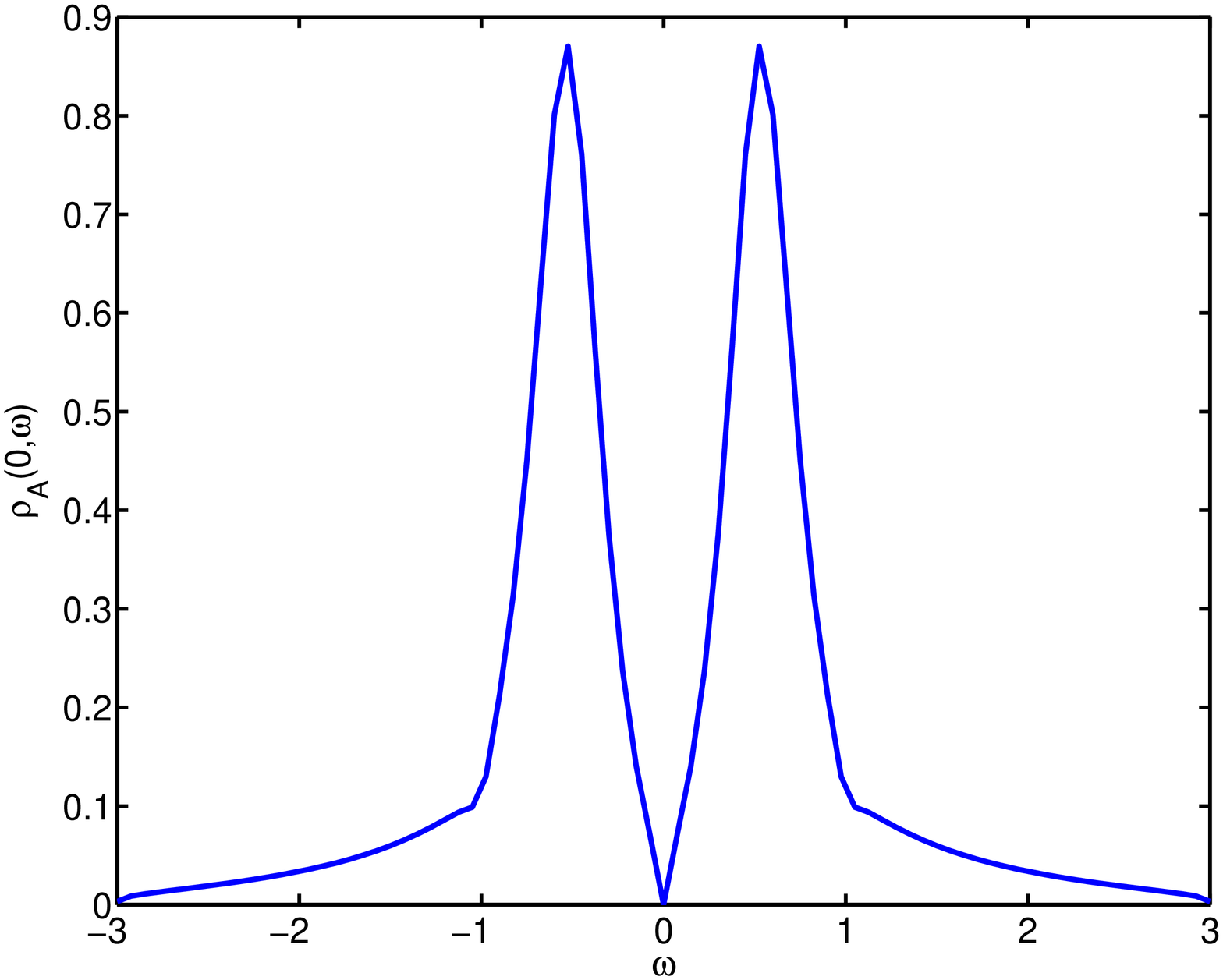}\includegraphics*[angle=-90,totalheight=3.5cm,width=4.5cm,
viewport=10 10 560 690,clip]{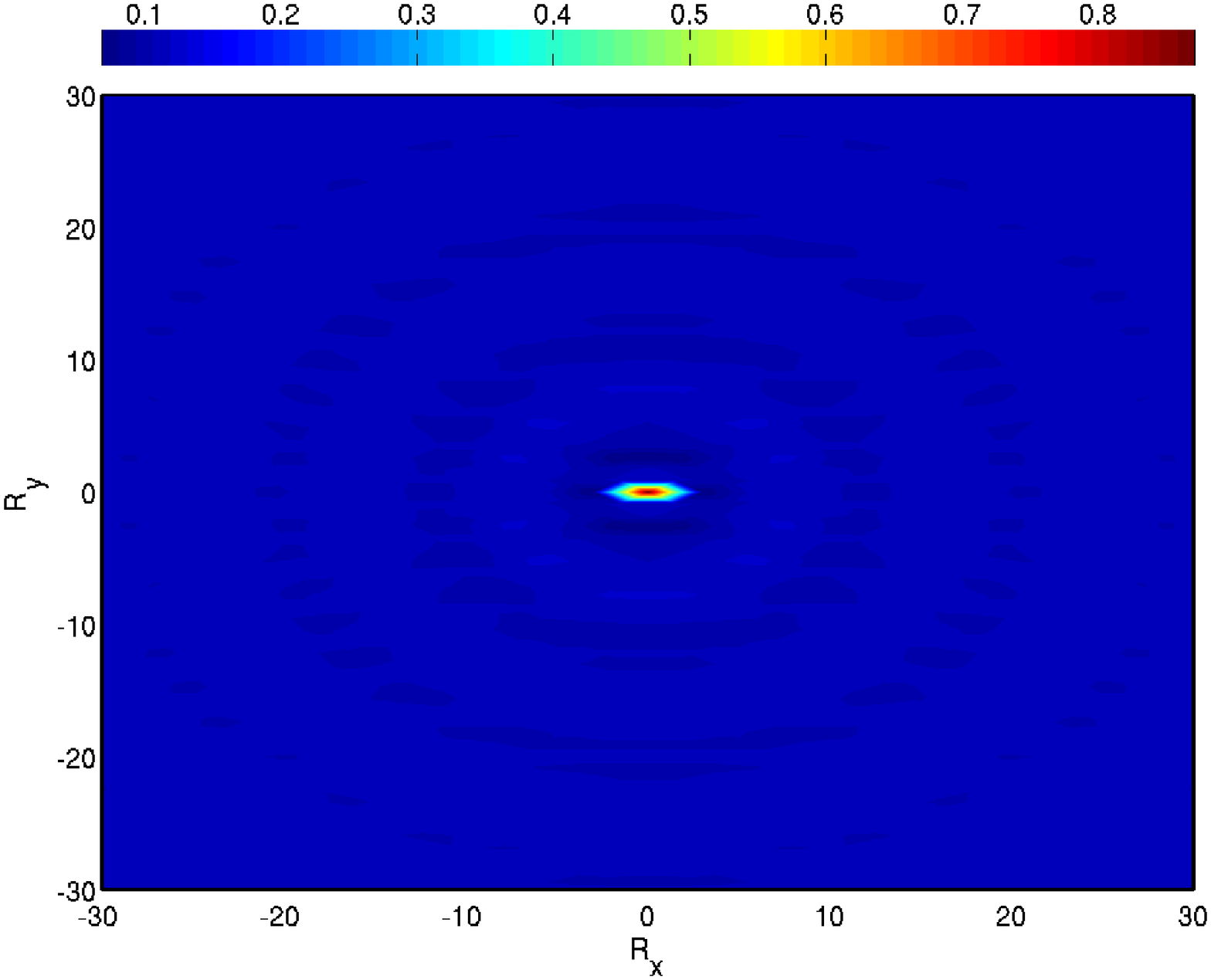}}
{\includegraphics*[angle=-90,totalheight=2.5cm,width=3.5cm,viewport=10 5 555 705,
clip]{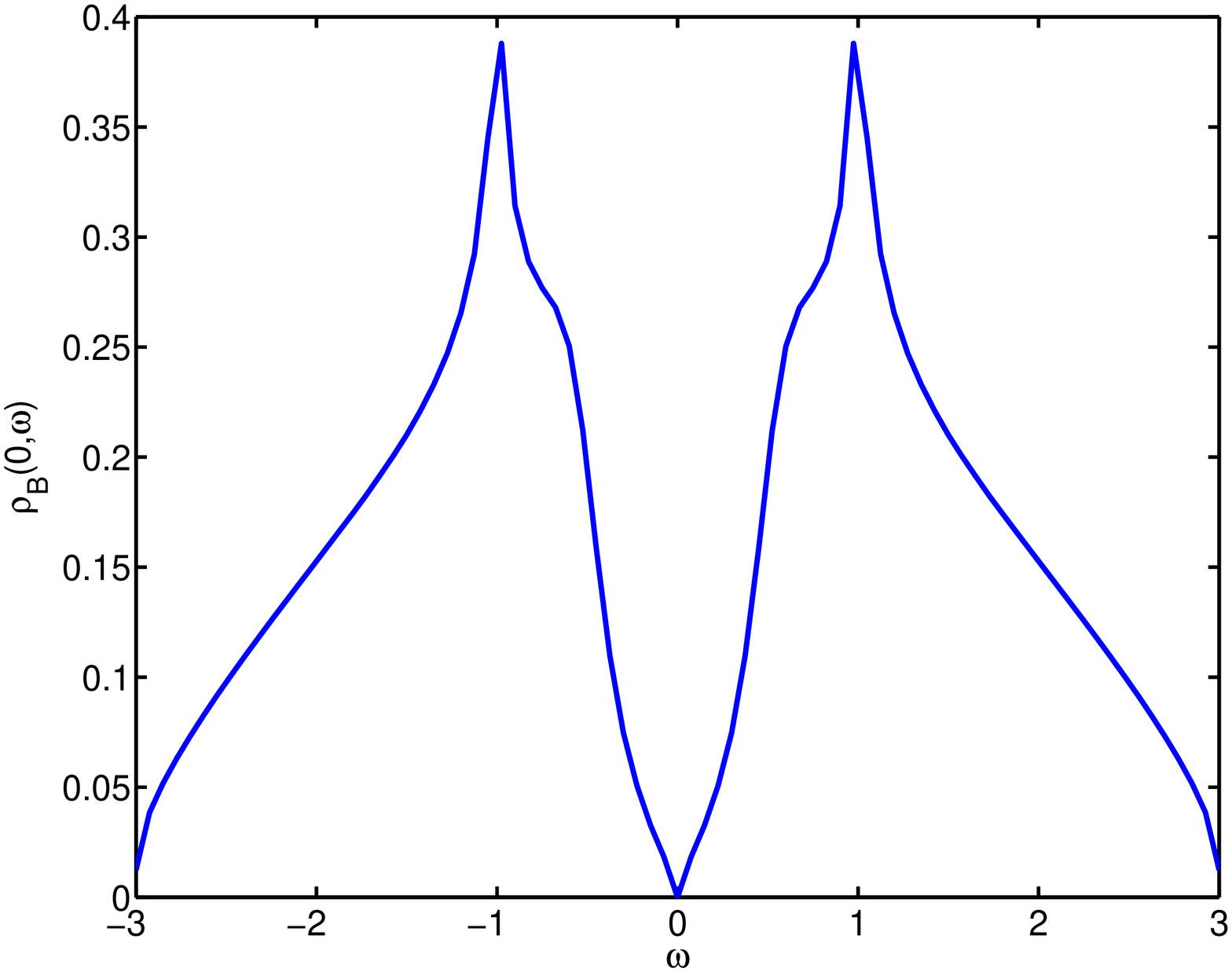}\includegraphics*[angle=-90,totalheight=3.5cm,width=4.5cm,
viewport=10 10 560 690,clip]{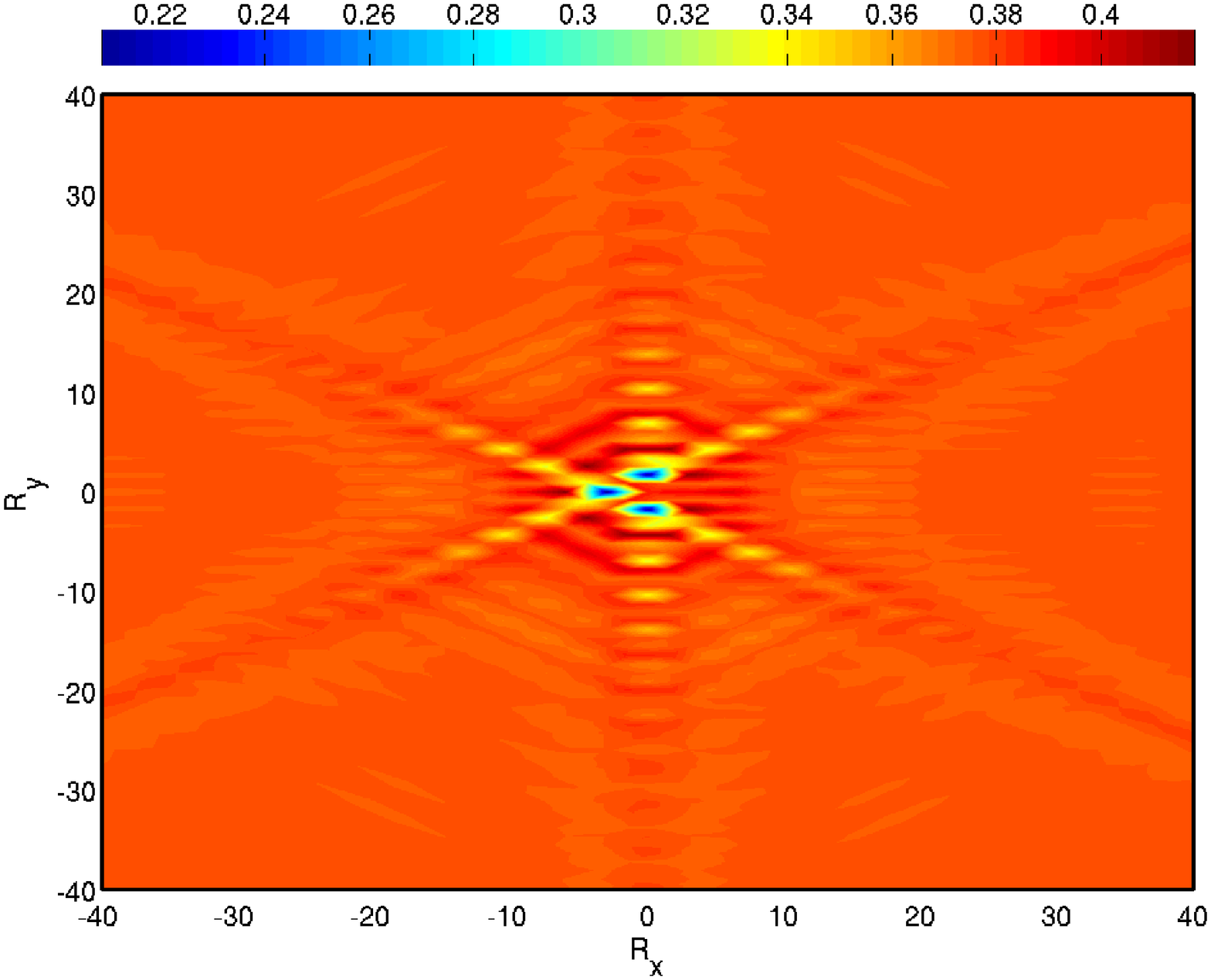}}
\end{center}
\caption{(color on line) Left: LDOS at the impurity site (above) and at its
nearest neighbors (below), for $t_0=0.5 t$. 
Right: intensity plots of $\rho_a(\bm r,\omega=-0.525t)$ (top), and 
$\rho_b(\bm r,\omega=-0.975t)$ (bottom).}
\label{fig:rrw05}
\end{figure}

We first study the effects of hopping disorder alone, 
and consider the case $t_0=0.5t$. Fig.~(\ref{fig:nr05}) shows an intensity plot of the 
real-space distribution of the number of electrons, given by
(\ref{friedel}). Interference effects give rise to Friedel oscillations displaying 
the underlying six-fold symmetry. The main contributions in the sum over negative 
states (\ref{friedel}) come from: 
(i) the resonance states created by the impurity; and (ii) the spectral weight
under the van Hove singularity that is also affected by the impurity.   
Fig.~(\ref{fig:rrw05}) shows these two contributions separately.  In the left,  we show the LDOS spectrum at the impurity site, and at its 
nearest-neighbors (B-sites). The spectrum shows a resonance peak at 
$\omega_0=\pm 0.525t$ on the impurity, while for its nearest-neighbors it is 
mostly dominated by the van Hove peaks at $\omega\approx \pm t$ 
($\omega_0=\pm 0.975t$). The size of these peaks decreases away 
from the impurity and Fig.~(\ref{fig:rrw05}) shows the
intensity plots of these peaks in real-space, which can be directly measured
using STM spectroscopy. 

\begin{figure}[thf]
\begin{center}
{\includegraphics*[angle=-90,totalheight=3.0cm,width=4.0cm,viewport=10 5 555 695,
clip]{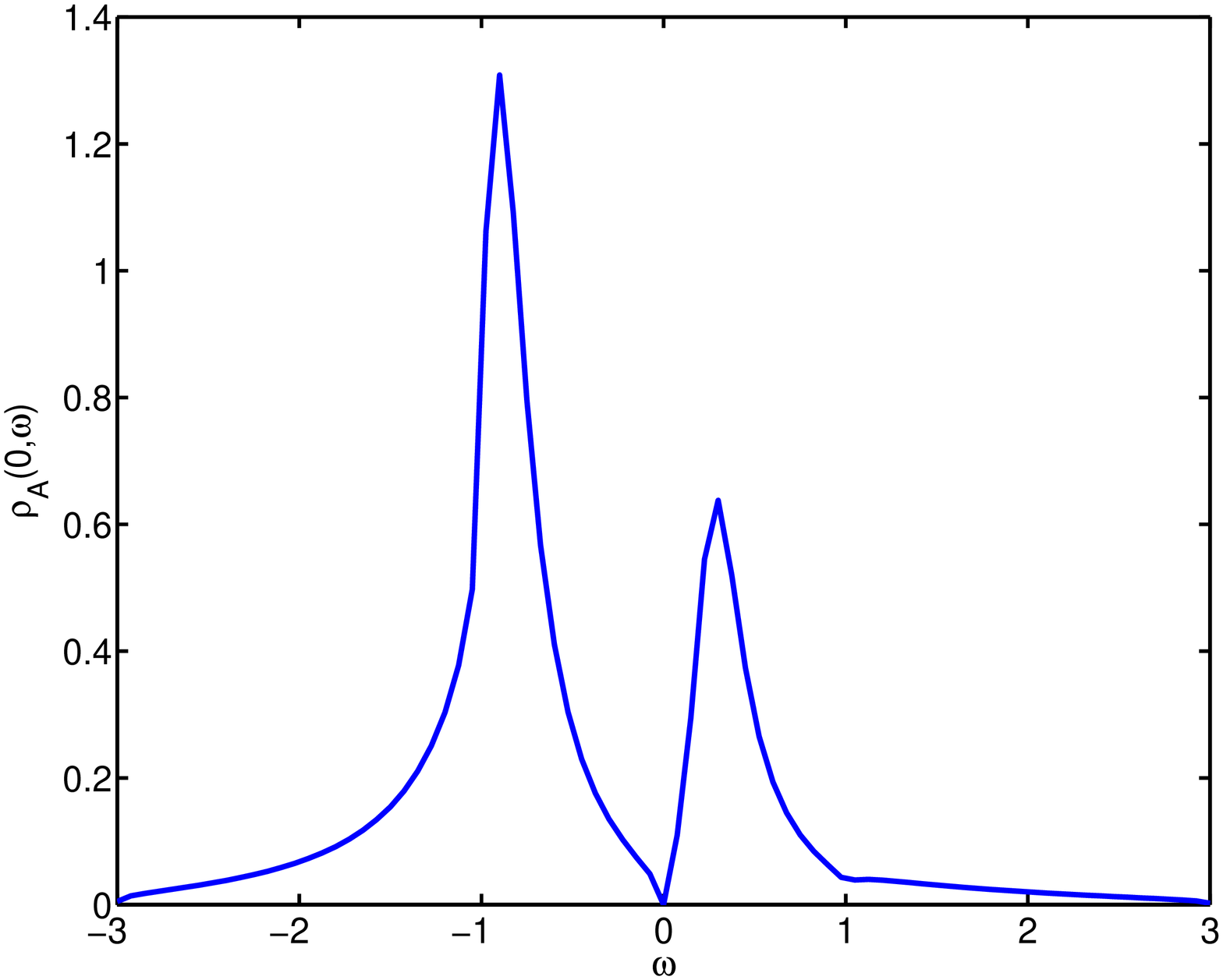}\includegraphics*[angle=-90,totalheight=3.0cm,width=4.0cm,
viewport=10 5 555 720,clip]{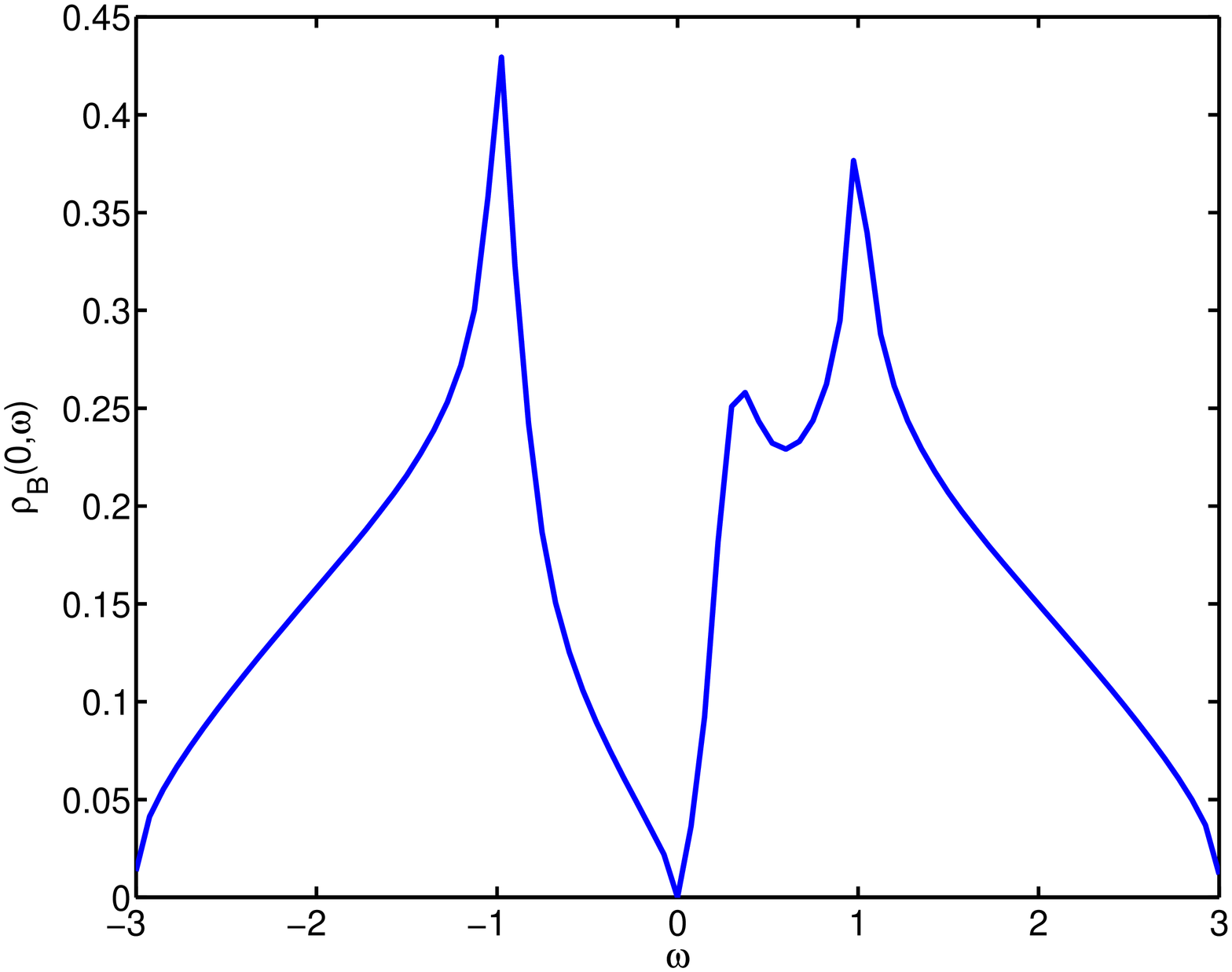}}
\end{center}
\caption{LDOS for the case $t_0=0.5t$ and $\varepsilon_0=-0.525t$, at the impurity site (left) and at the nearest neighbor sites (right).}
\label{fig:nitrogen}
\end{figure}

When a negative $\varepsilon_0$ is also added, the LDOS gets modified as shown in 
Fig.~(\ref{fig:nitrogen}) for the impurity and nearest neighbor sites. 
Any finite potential value at the impurity breaks particle-hole symmetry and as
a result the spectrum becomes asymmetric under reflection 
($\omega\rightarrow -\omega$).

\begin{figure}[htf]
\begin{center}
{\includegraphics*[angle=-90,totalheight=2.5cm,width=3.5cm,viewport=5 5 560 705,
clip]{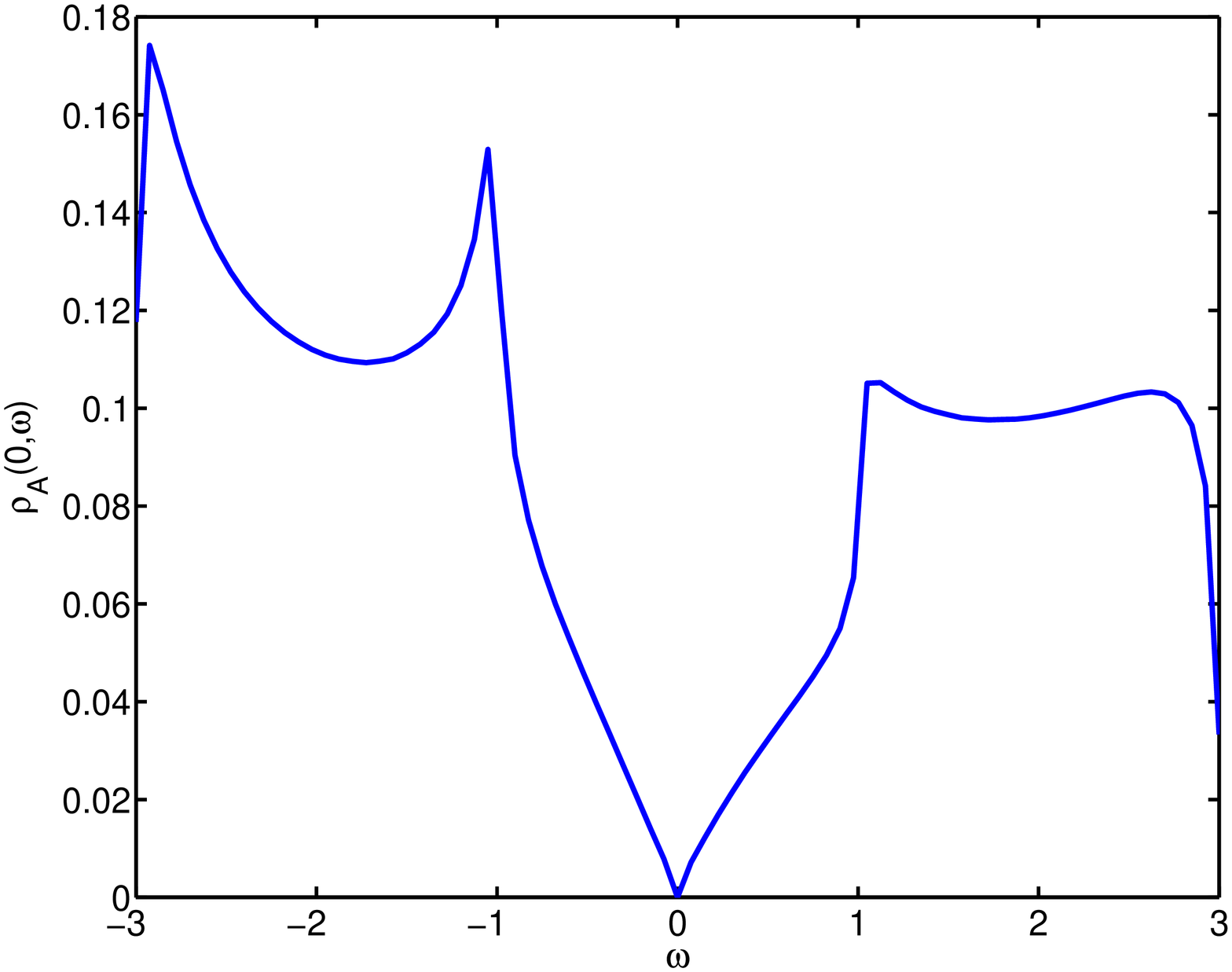}\includegraphics*[angle=-90,totalheight=3.5cm,width=4.5cm,
viewport=5 10 560 690,clip]{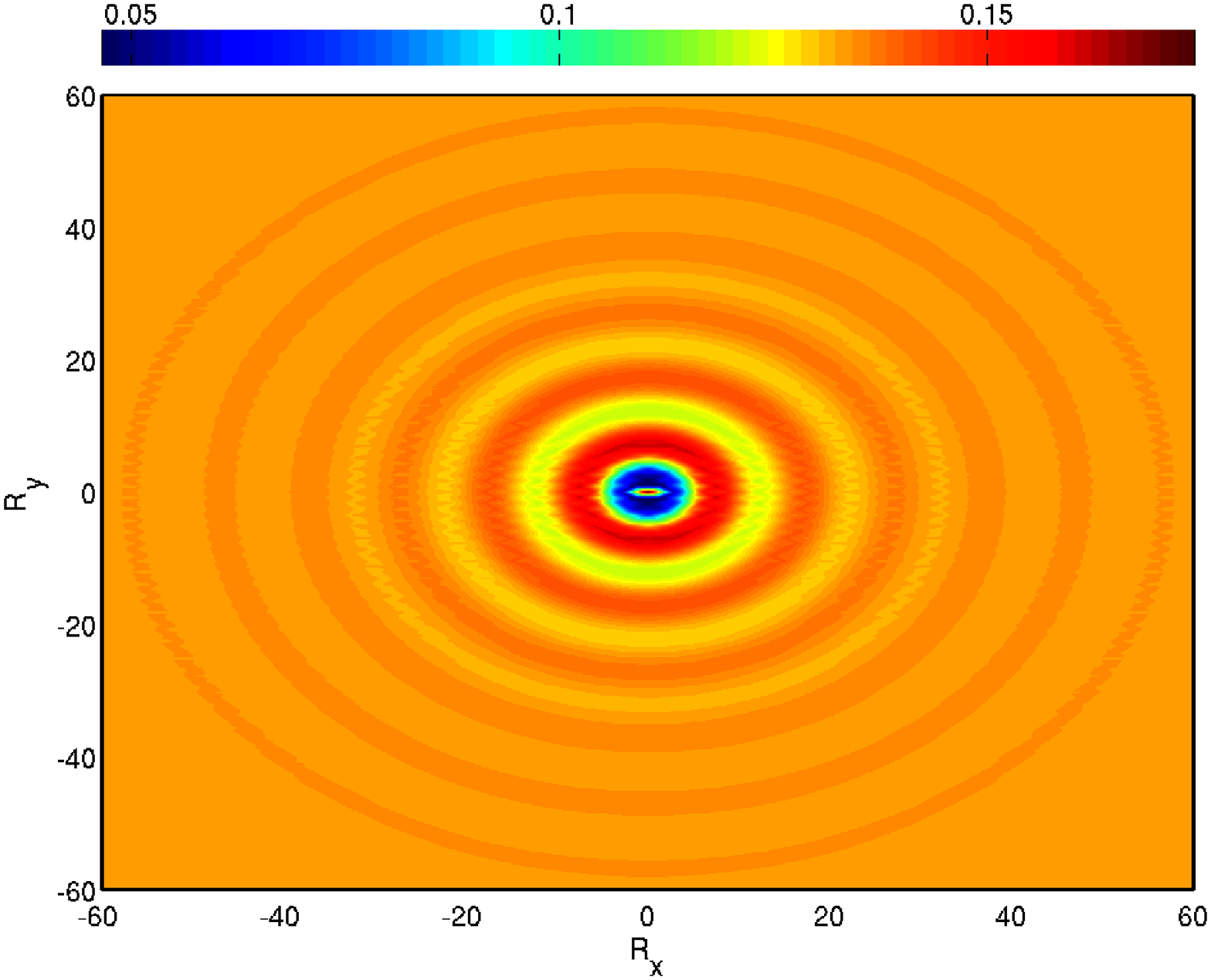}}
{\includegraphics*[angle=-90,totalheight=2.5cm,width=3.5cm,viewport=5 5 560 705,
clip]{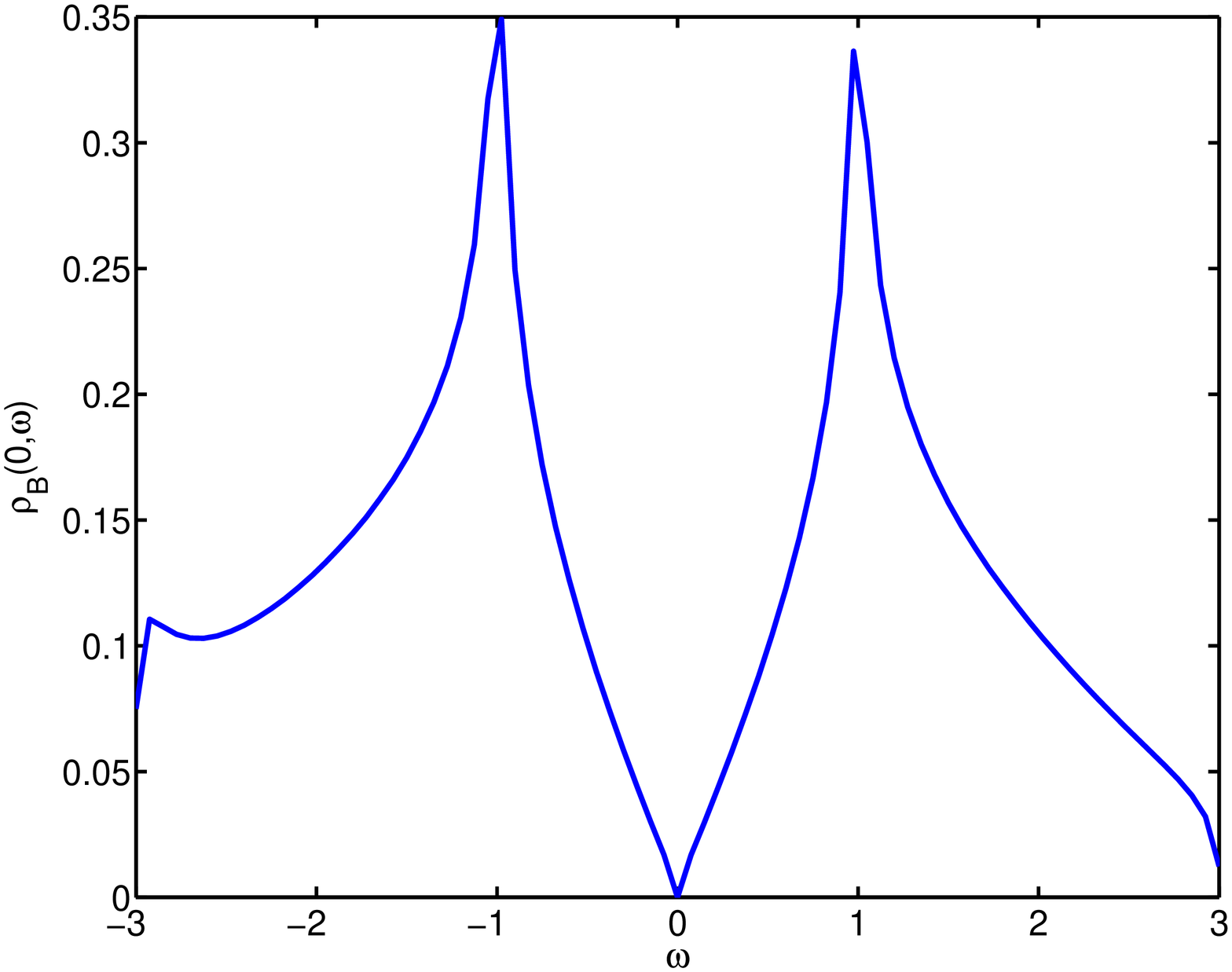}\includegraphics*[angle=-90,totalheight=3.5cm,width=4.5cm,
viewport=5 10 560 690,clip]{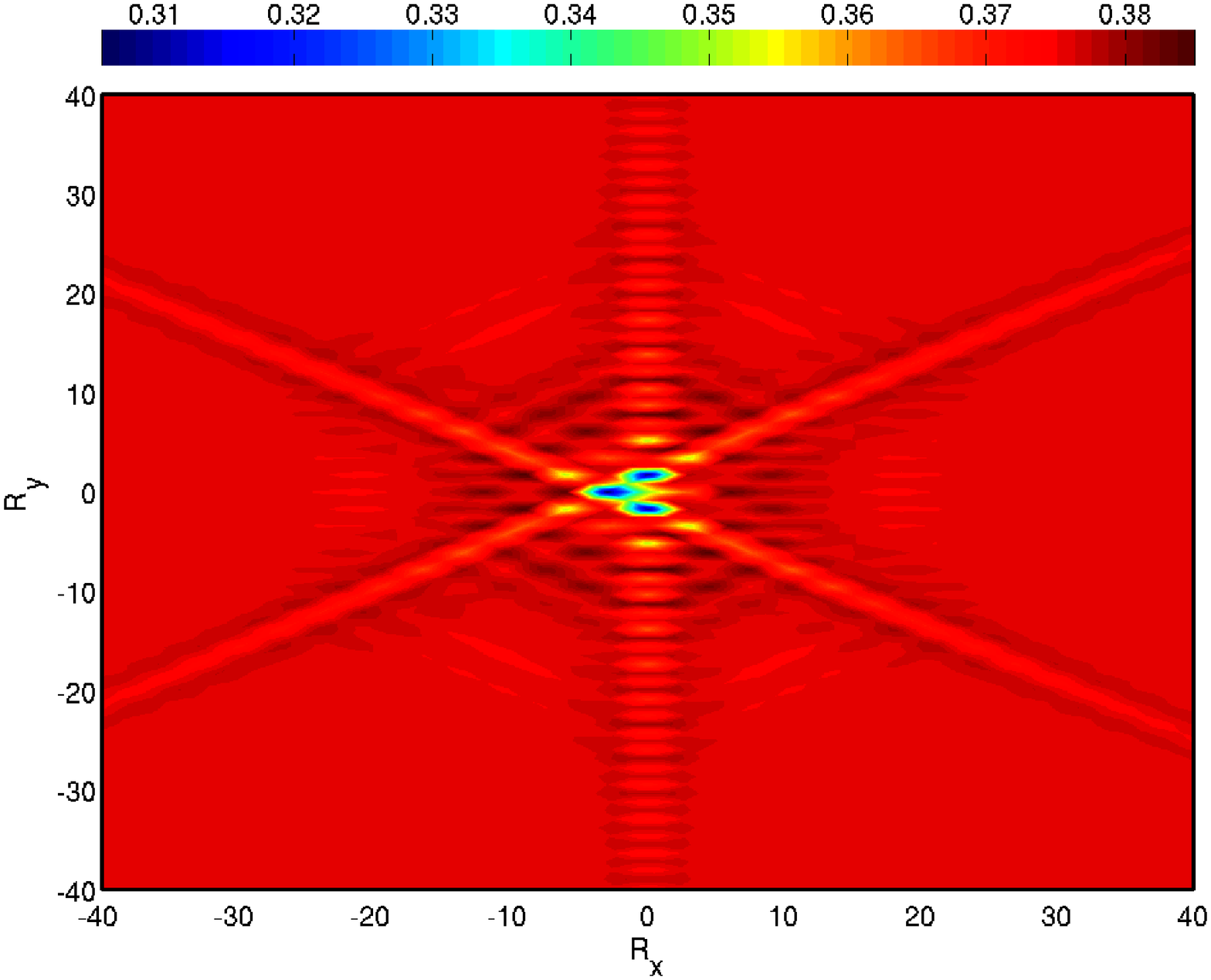}}
\end{center}
\caption{(color on line) Electronic spectra for the case $t_0=-0.5t$ and 
$\varepsilon_0=0.525t$, corresponding to an impurity such as boron. 
Left: LDOS at the impurity site (top) and nearest neighbors (bottom). 
Right: real-space intensity plot of $\rho_a(\bm r,\omega=-2.935t)$ (top), and
$\rho_b(\bm r,\omega=-0.975t)$ (bottom).}
\label{fig:boron}
\end{figure}
{\it Boron substitution.} In this case the impurity site has larger hopping 
amplitude and a positive on-site potential. Fig.~(\ref{fig:boron}) shows the 
LDOS at the impurity and the nearest neighbor sites for a case with $t_0=-0.5t$
(so that the hopping amplitude to the impurity, $t-t_0$, is larger than the
homogeneous system) and $\varepsilon_0=0.525t$. 
The peaks of the LDOS at certain energies originate from impurity resonance
states and van Hove singularities. The real-space intensity plots are at
resonance: $\omega_0=-2.935t$ for the sublattice-A LDOS, and $\omega_0=-0.975t$
(at the van Hove singularity) for the sublattice-B LDOS. 
Just as for the case of N shown in Fig.~(\ref{fig:rrw05}) the van Hove
singularities are more strongly affected at certain directions. The van Hove
peaks are strongest on B-sites where the LDOS can be understood as being
originated by three impurities (the three nearest neighbor B-sites of the actual
impurity site) and hence the star-shaped symmetry shown in the intensity maps at
$\omega_0=-0.975t$ in Fig.~(\ref{fig:rrw05}) and Fig.~(\ref{fig:boron}). 
Near the band edge, the band structure of the clean system behave more like a conventional 2D system, so the resonance peak at $\omega_0=-2.935t$ has a symmetric spatial distribution (Fig. (\ref{fig:boron}), top).

{\it Conclusions.} We find the exact electronic Green's functions 
in the presence of an impurity that modifies both the local 
atomic energy as well as the hopping amplitude with neighboring
C atoms. We have shown that the presence of the impurity leads to
strong modifications of the local density of states and also to
the presence of unusual Friedel oscillations, both quantities being
accessible to measurement through STM spectroscopy. We have applied
our theory to the case of substitutional B and N and found that 
these two atoms have very specific spectroscopic signatures.
From the knowledge of these Green's function one can also calculate
the transport properties of chemically substituted graphene. 

\acknowledgments
We thank the hospitality of the KITP at UCSB,
(NSF grant No. PHY05-51164), 
where part of this work was performed. 
We thank A. D. Klironomos and V. M. Pereira for useful 
discussions.
JMBLS
and NMRP acknowledge financial support from POCI 2010 via project
PTDC/FIS/64404/2006. A.H.C.N. was supported through NSF grant DMR-0343790.


\end{document}